\documentclass{article}
\usepackage{spconf,amsmath,graphicx}
\usepackage{multirow}
\usepackage{booktabs}
\usepackage{makecell}
\newcommand{\specialcell}[2][c]{%
  \begin{tabular}[#1]{@{}l@{}}#2\end{tabular}}
\usepackage{amsmath}
\DeclareMathOperator*{\argmax}{argmax}
\usepackage{color}
\newcommand{\then}{$\rightarrow\,$}

\title{A spelling correction model for end-to-end speech recognition} 
\name{Jinxi Guo$^{1*}$ \thanks{*Work done as an intern at Google NYC}, Tara N. Sainath$^2$, Ron J. Weiss$^2$}
\address{$^1$University of California, Los Angeles, USA \\ $^2$Google Inc., USA\\
\texttt{\normalsize lennyguo@g.ucla.edu, \{tsainath, ronw\}@google.com}}

\begin{document}
\ninept
\maketitle

\begin{abstract}
Attention-based sequence-to-sequence models for speech recognition jointly train an acoustic model, language model (LM), and alignment mechanism using a single neural network and require only parallel audio-text pairs. Thus, the language model component of the end-to-end model is only trained on transcribed audio-text pairs, which leads to performance degradation especially on rare words. While there have been a variety of work that look at incorporating an external LM trained on text-only data into the end-to-end framework, none of them have taken into account the characteristic error distribution made by the model. In this paper, we propose a novel approach to utilizing text-only data, by training a spelling correction (SC) model to explicitly correct those errors. On the LibriSpeech dataset, we demonstrate that the proposed model results in an 18.6\% relative improvement in WER over the baseline model when directly correcting top ASR hypothesis, and a 29.0\% relative improvement when further rescoring an expanded n-best list using an external LM.
\end{abstract}
\section{Introduction}
\label{sec:intro}
End-to-end models for automatic speech recognition (ASR) have gained popularity in recent years as a way to fold separate components of a conventional ASR system (i.e., acoustic, pronunciation and language models) into a single neural network \cite{Chan15,Bahdanau16,Graves12,Graves06}. The Listen, Attend and Spell (LAS) model \cite{Chan15}, is one such model which has shown competitive performance on a Voice search task compared to a strong conventional baseline model \cite{CC18}.

End-to-end models require audio-text pairs during training. They are therefore trained using far less data compared to the language model (LM) component of a conventional recognizer, which is often trained with an order of magnitude more text-only data. Due this reduced training data, end-to-end models do not perform as well on utterances containing rare words which occur infrequently in the audio-text training set.
To address this issue, there have been a variety of research studies that look at incorporating an RNN-LM trained on text-only data into the end-to-end framework. These approaches include rescoring the n-best decoded hypotheses from the end-to-end model \cite{Chan15}, or incorporating an RNN-LM into the first-pass beam search \cite{Bahdanau16, Jan17, Anjuli18, Toshniwal18}, via shallow, cold or deep fusion. 

While such RNN-LM fusion techniques fix some tail words cases, we have found that numerous rare word and proper noun errors still exist. One hypothesis is that the LM is not integrated into the end-to-end model with the objective of correcting errors that the end-to-end model makes. Specifically, the LM is either (a) fused into the end-to-end model to minimize some loss on the training data (deep, cold fusion)  or (b) incorporated externally during decoding without any training objective (rescoring, shallow fusion).

The goal of this paper is to incorporate a module trained on text-only data into the end-to-end framework, with the objective of correcting errors made by the system. Specifically, we investigate using unpaired text-only data to synthetically generate corresponding audio signals using a text-to-speech (TTS) system, a process similar to backtranslation in machine translation \cite{Sennrich16}. We then run the baseline LAS speech recognizer on the TTS output to create a set of text-to-text pairs representing an error hypothesis and its corresponding ground truth. We train a spelling corrector (SC) model on these text-to-text pairs, to correct potential errors made by the first-pass recognizer. We compare our proposed approach to two other approaches to incorporating text-only data which do not account for the error distribution of the LAS model: rescoring a n-best list with an RNN-LM,  and directly training the LAS model on TTS-synthesized speech. 

For related work, \cite{Errattahi18} contains an overview of previous  work on error correction for ASR. Only few of them could detect and correct the errors automatically. \cite{Sarma04} built an ASR error detector and corrector using co-occurrence analysis. \cite{Assil12} proposed post-editing ASR errors based on an external N-gram dataset. More recently, \cite{Xie16} proposed an attention-based network architecture similar to ours for a language correction task.

We evaluate our proposed approach on a 960-hour LibriSpeech task. Our results show that the proposed spelling correction models results in an 18.6\% relative improvement in WER when directly correcting the top hypothesis emitted by the LAS model, and a 29.0\% relative improvement when further rescoring an expanded n-best list generated by the correction model using an external LM, demonstrating the increased diversity introduced by the SC model, and significantly outperforming simple LM rescoring and direct LAS model training on the TTS data.

\section{Baseline recognition model}
\label{sec:las baseline}
The baseline speech recognition model used for experiments in this paper is a LAS-inspired encoder-decoder architecture with attention based on \cite{Yu17}.
The encoder consists of a stack of covolutional and LSTM layers, which takes input mel spectrogram features, $\mathbf{x}$, and maps it to a higher order feature representation $\mathbf{h}^{enc}$.
The encoder output is passed to an attention mechanism, which aligns the input audio sequence with the output sequence representing the transcript, determining which encoder frames should be used to predict the next output symbol, $y_i$. The output of the attention is a context vector $\mathbf{c_i}$ that is passed to the decoder.
Finally, the decoder takes the attention context $\mathbf{c_i}$ as well as an embedding of the previous prediction, $y_{i-1}$, and generates logits $\mathbf{h}^{dec}$. These logits are passed through a softmax to compute a probability distribution $P(y_i|\mathbf{h}^{dec})$ across output tokens. We can think of the decoder as being similar to a language model. The model is trained to minimize the cross-entropy loss on the training data. In our work, the decoder outputs a sequence of wordpiece units $y_i$, which has shown good performance for both ASR and machine translation (MT) tasks \cite{CC18,Yonghui16,Zeyer18}.

\section{Utilizing text-only data}
\label{sec:uti-text-data}
\subsection{External LM}
\label{sec:ext-lm}
A common approach for incorporating text-only data is to train an RNN-LM on text-data \cite{Chan15, Bahdanau16, Jan17, Anjuli18, Toshniwal18}, and then incorporating the LM during beam search decoding through various mechanisms proposed in the literature. For example, \cite{Chan15} demonstrated significant improvement by simply using an LM to rescore the n-best hypotheses produced by LAS. \cite{Bahdanau16} extended this idea by performing log-linear interpolation between LAS and an LM at each step of the beam search, a method referred to as shallow fusion. Shallow fusion was further studied in \cite{Jan17, Anjuli18}. Gulcehre, et al~\cite{Gulcehre15} proposed a deep fusion method, which integrates the external LM into encoder-decoder by fusing together the hidden states. Cold fusion, a modification of deep fusion by doing early training integration was proposed in \cite{Sriram17}. Both deep and cold fusion methods are further investigated for ASR in \cite{Toshniwal18}.

In this paper we focus on n-best rescoring similar to \cite{Chan15}. LM rescoring makes it easier to evaluate the spelling corrector which we will introduce in Section \ref{sssec:inc_nbest}, and thus for a fair comparison we did the same for the baseline LAS model. Specifically, during inference our objective is to find the most likely sub-word unit sequence given the score from the LAS model $P(\mathbf{y}|\mathbf{x})$ and the LM $P_{LM}(\mathbf{y})$:
\begin{equation}
\mathbf{y}^{*} = \argmax_\mathbf{y} \log P(\mathbf{y}|\mathbf{x}) + \lambda \log P_{LM}(\mathbf{y})
\label{eq:las+lm}
\end{equation}
where $\lambda$ is an interpolation weight determined on a held-out set.

\subsection{Training on synthesized speech}
\label{sec:train-with-tts}
Another approach is to use text-only data to synthesize audio-text training data using a text-to-speech (TTS) system.  An analogous approach has been explored in machine translation, where Sennrich \cite{Sennrich16} passed unparied text in the target language into a pretrained ``backtranslation'' model in order to generate corresponding text in the source language.   This synthetic data was then used to augment the existing parallel training data to train the translation model. In this work we synthesize the text-only data using a high quality TTS system based on parallel WaveNet \cite{Aaron18} and use the resulting synthetic audio when training the LAS model. A similar approach was recently applied to speech recognition in \cite{Hayashi18}.

\subsection{Spelling correction model}
\label{sec:sc-model}
A disadvantage of the previously described approaches for incorporating text-only training data is that they do not take into account the characteristic error distribution made by the speech recognizer. In this section we propose a novel approach to utilizing text-only data, by training a supervised ``spelling correction'' model to explicitly correct the errors made by the LAS recognizer.
Intuitively, this task is simpler than unsupervised language modeling because it is able to take advantage of the existing language modeling capacity of the LAS model.  Instead of predicting the likelihood of emitting a word based on the surrounding context as in an RNN-LM, the SC model instead needs only to identify likely errors in the LAS output and propose alternatives.  Since the baseline LAS model already has a relatively low word error rate, most of the time this task reduces to simply copying the input transcript directly to the output.

\subsubsection{Training data}
Training an SC model requires a parallel text training set consisting of LAS hypotheses to be corrected and ground truth text sequences.
In order to generate a training corpus from text-only data which is representative of the baseline LAS model's error distribution we generate TTS utterances \{$u_1, u_2, ...$ \} using text sequences \{$y_1, y_2, ...$\} from the text corpus. Next we perform decoding on the TTS data using the pretrained LAS model. Each TTS utterance $u_i$ can be paired with $N$ hypotheses \{$H_{i1}, H_{i2}, ..., H_{iN}$\} after beam-search decoding. By using all hypotheses from the n-best list to generate training data, we create a diverse training set that captures more variance of the LAS model's underlying error distribution. During SC training we randomly sample a hypothesis $H_{ij}$ from the LAS n-best list and combine it with the ground-truth transcript $y_{i}$ to form a training pair.

\subsubsection{Architecture and training}
We use an attention-based encoder-decoder sequence-to-sequence architecture for our spelling corrector, similar to the neural machine translation model from \cite{Mia18}.
The input and output sentences are first decoded as wordpieces \cite{Schuster2012}. The encoder takes the embedding learned from the input sequence $H_{ij}$ and maps it to a higher-level representation through a stack of bi-directional LSTM layers. The decoder also consists of stacked unidirectional LSTM layers that uses an attention mechanism to attend to the encoder representation and generate the output sequence $y_i$ one token at a time. 

\begin{figure}[t]
\centering
\includegraphics[width=1.0\linewidth]{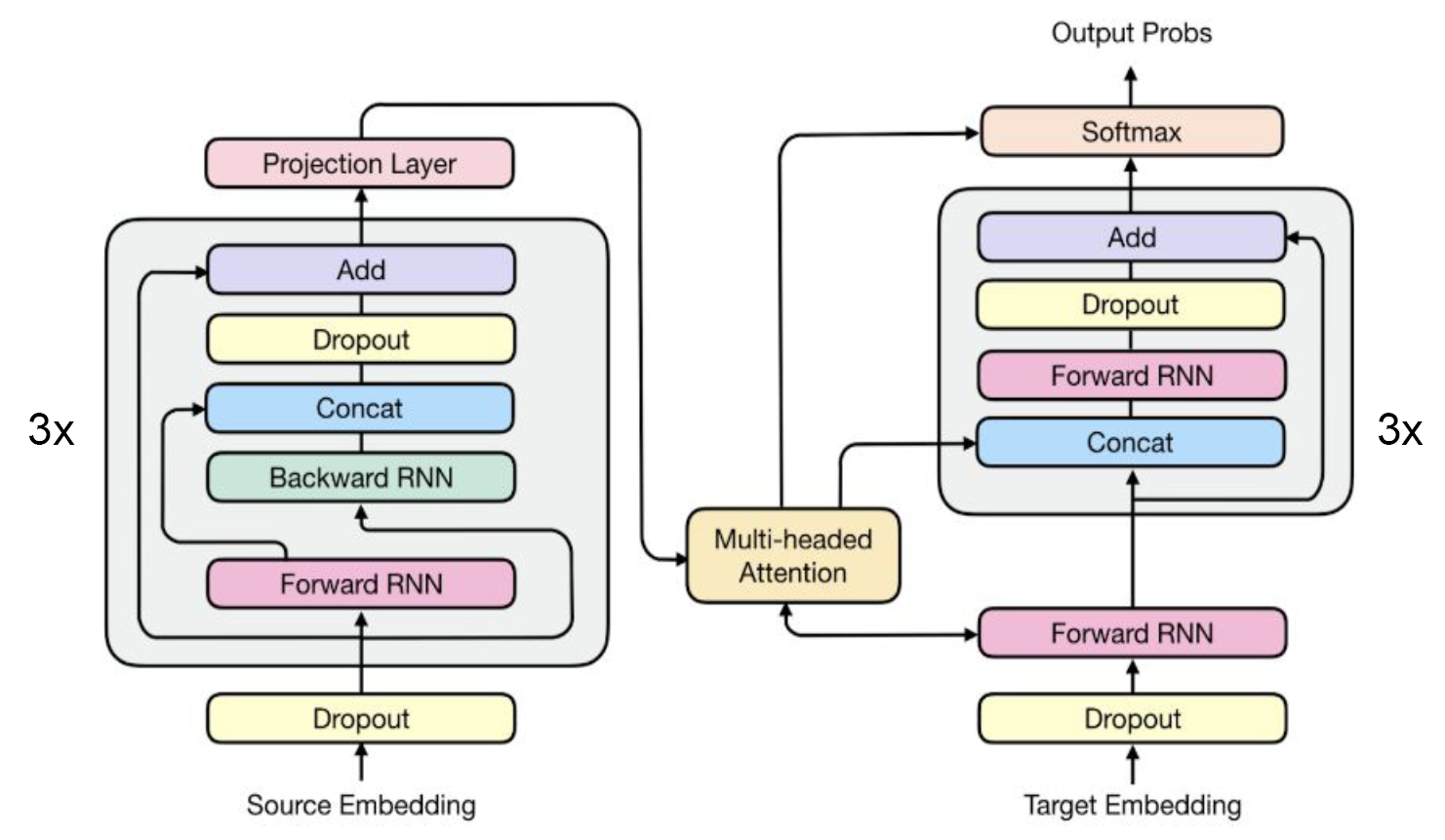}
\vspace{-6.5mm}
\caption{Spelling Correction model architecture.}
\label{fig:sc_model}
\vspace{-4.5mm}
\end{figure}

Figure~\ref{fig:sc_model} shows the model architecture. Compared with the standard attention-based sequence-to-sequence model, there are several differences. First, residual connections are added between layers in both the encoder and decoder. Per-gate layer normalization is applied within each LSTM cell, and multi-head additive attention is used. 
The bottom decoder layer and the final encoder layer output are used to obtain the recurrent attention context,which 
is fed to all decoder LSTM layers and also the softmax layer by concatenation.
The model is trained using the standard maximum-likelihood criterion, to maximize the sum of log probabilities of the ground-truth outputs given the corresponding inputs.

\subsubsection{Inference}
\label{sssec:inc_nbest}

During inference, decoding the LAS model with beam serach takes an utterance $u$ and produces $N$ hypotheses \{$H_1, H_2, \ldots, H_N$\} with corrsponding log probability scores \{$p_1, p_2, \ldots, p_N$\}. For hypothesis $H_i$, the SC model can similarly be used to generate $M$ new hypotheses \{$A_{i1}, A_{i2}, \ldots, A_{iM}$\} with corresponding log probability scores \{$q_{i1}, q_{i2}, \ldots, q_{iM}$\}.
Therefore, for a given utterance $u$, the cascaded LAS and SC models can generate a total of $N \times M$ hypotheses with associated scores.
Rescoring all $N \times M$ candidates with an LM gives a set of LM scores \{$r_{11}, r_{12}, \ldots, r_{MN}$\}. Finally, we can find the most likely hypothesis using the following criteria:
\begin{equation}
\mathbf{A}^{*} = \argmax_\mathbf{A} \lambda_{LAS}*p_i + \lambda_{SC}*q_{ij} + \lambda_{LM}*r_{ij}
\label{eq:las+sc+lm}
\end{equation}
where $\lambda_{LAS}$, $\lambda_{SC}$ and $\lambda_{LM}$ are the weights for LAS, SC, and LM scores respectively, which are determined on a held-out set.
Note that independently correcting each of the $N$ LAS hypotheses with the SC model increases the total computational cost of the system by a factor of $N$.  In the following experiment section we compare configurations where the SC model is used to correct only the top LAS hypothesis, i.e.\ using $N=1$, to the full $N\times M$ configuration.

\section{Experimental setup}
\label{sec:expsetup}

We conduct our experiments on LibriSpeech \cite{Libri15}. The training data contains around 960 hours of speech from read audio book recordings. We evaluate on the ``clean'' dev and test sets, which each contains around 5.4 hours of speech. For feature extraction, 80 dimensional log-mel filterbank features are computed from 25ms windows shifted by 10ms, stacked with their deltas and accelerations.

As an external text-only training dataset, we use the 800M word LibriSpeech language modeling corpus which was carefully selected to avoid overlap with the text in the dev and test sets. We preproccess the corpus by filtering out 0.5M sequences which contain only single letter words or are longer than 90 words.  We use the remaining 40M text sequences to train the language model, generate a TTS dataset to train the LAS model, and generate error hypotheses to train the SC model. 

\subsection{Speech recognition model}
\label{sssec:subsubhead}
The baseline LAS recognition model uses 2 convolutional layers and 3 bidirectional LSTM layers in the encoder, and a single undirectional LSTM layer in the decoder. A multi-headed additive attention mechanism with 4 heads is used. We use a wordpiece model with 16K tokens, which is generated using the byte pair encoding algorithm. The wordpieces are represented with 96 dimensional embeddings which are learned jointly with the rest of the model. For regularization, label smoothing \cite{Jan17} is applied by weighing the ground truth token at each output step by 0.9, and uniformly distributing the remaining probability mass among other tokens. During inference we use beam search decoding with a beam size of 8. 
The baseline LAS model is trained using asynchronous stochastic gradient descent optimization with 16 workers. All models are implemented in Tensorflow \cite{AbadiAgarwalBarhamEtAl15} and trained using the Adam optimizer \cite{KingmaBa15}.

\subsection{Language model}
\label{sssec:subsubhead}
We train a stacked RNN with two unidirectional LSTM layers to use as an external language model. The same 16K wordpiece token set is used as the LAS model.
Early stopping during training is based on the dev set perplexity.
Similar to \cite{Chan15}, we use the LM to rescore the n-best list generated by decoding the LAS model with beam search. The interpolation weight $\lambda$ is swept on a held-out dev set.

\subsection{TTS model}
\label{sssec:tts}
In order to synthesize speech, we use a Parallel WaveNet model \cite{Aaron18}, which can generate high fidelity speech very efficiently, trained on 65-hours of speech from a single female speaker. We use this model to perform inference on the full text-only dataset, and generate 40M audio utterances.
We train a combined real+TTS LAS recognizer by combining these synthetic speech utterances with the 960 hour LibriSpeech training set. During training, each batch is comprised of 70\% real speech and 30\% synthetic speech.

\subsection{Spelling correction model}
\label{sssec:subsubhead}
The proposed SC model uses 3 bidirectional LSTM layers in the encoder and 3 unidirectional LSTM layers in the decoder. Residual connections are added to the third layer of both the encoder and decoder. Four-headed additive attention is used. The same 16K wordpiece model is adopted as in the LAS and language models. Beam search decoding is performed with a beam size of 8.
For regularization, a dropout rate of 0.2 is applied to both embedding layers and each LSTM layer output. Attention dropout is also applied with the same rate. Uniform label smoothing with uncertainty 0.1 is applied, and parameters are L2 regularized with weight $10^{-5}$. 

The learning rate schedule includes an initial linear warm-up phase, a constant phase, and a exponential decay phase following \cite{Mia18}. SC model is trained with synchronous training using 32 GPUs and adaptive gradient clipping is used to further stabilize training.

To generate training data for this model, we first decode the 40M clean TTS utterances using the baseline LAS model. Each TTS utterance results in 8 hypotheses, and each hypothesis is grouped with the corresponding ground-truth transcript to form an SC training pair. This process results in a total of about 320M training pairs.

We additionally experiment with a multi-style training (MTR) configuration using an augmented dataset by corrupting the synthesized utterances with additive noise and reverberation using a room simulator.
Noise signals are collected from YouTube and daily life noisy environmental recordings, and randomly reverberated and mixed with the speech such
that the overall SNR is between 20dB and 40dB\cite{Chanwoo17}. We run the full synthesized speech set through this process, resulting in a total of 40M noisy utterances.  These utterances are decoded by the baseline LAS model following the same procedure described above to generate the noisy SC training set.  The union of clean and noisy train sets yields a total of 640M MTR pairs.

\section{Results}
\label{sec:majhead}

In this section, we will compare different methods of incorporating text-only data.

Baseline results found using the LAS model with and without rescoring the n-best list using an external LM are shown in the top of Table~\ref{table:baseline-text-only} (top two rows). The optimal LM weight of $\lambda=0.5$ was selected by tuning on the development set.  LM rescoring leads to a relative improvement of 21.7\% over the LAS baseline.      

Next we present results by augmenting the LAS training set using speech synthesized from the text-only training data. As mentioned in Section~\ref{sssec:tts}, in order to avoid overfitting we train the LAS model using a combination of real and TTS audio, referred to as LAS-TTS. Table~\ref{table:baseline-text-only} shows that this improves performance slightly, but the gains are not as large as LM rescoring the baseline. However, the combination of TTS-augmented training and LM rescoring improves over LM rescoring along, demonstrating the complementarity of the two methods.

\begin{table}[t]
  \centering
  \begin{tabular}{lcc}
    \toprule
    \textbf{System} & \hspace{-0.5ex}\textbf{Dev-clean}\hspace{-0.5ex} & \hspace{-0.5ex}\textbf{Test-clean}\hspace{-0.5ex} \\
    \midrule
    LAS 		        & {5.80} & {6.03} \\
    LAS \then LM (8)           & 4.56 & 4.72 \\
    LAS-TTS 
                                        & 5.68 & 5.85 \\
    LAS-TTS \then LM (8) & \textbf{4.45} & \textbf{4.52} \\
    \midrule
    LAS \then SC (1)                    & 5.04 & 5.08 \\
    LAS \then SC (8) \then LM (64)      & \textbf{4.20} & \textbf{4.33} \\
    \midrule
    LAS \then SC-MTR (1)                & 4.87 & 4.91 \\
    LAS \then SC-MTR (8) \then LM (64)  & \textbf{4.12} & \textbf{4.28} \\
    \bottomrule
  \end{tabular}
  \vspace{-1.0mm}
  \caption{Word error rates (WERs) on LibriSpeech ``clean'' sets comparing different techniques for incorporating text-only training data.
    Numbers in parentheses indicate the number of input hypotheses considered by the corresponding model.
  }
  \label{table:baseline-text-only}
\end{table}

\subsection{SC model}

\begin{figure}[t]
\begin{center}
\centerline{\includegraphics[width=0.7\linewidth]{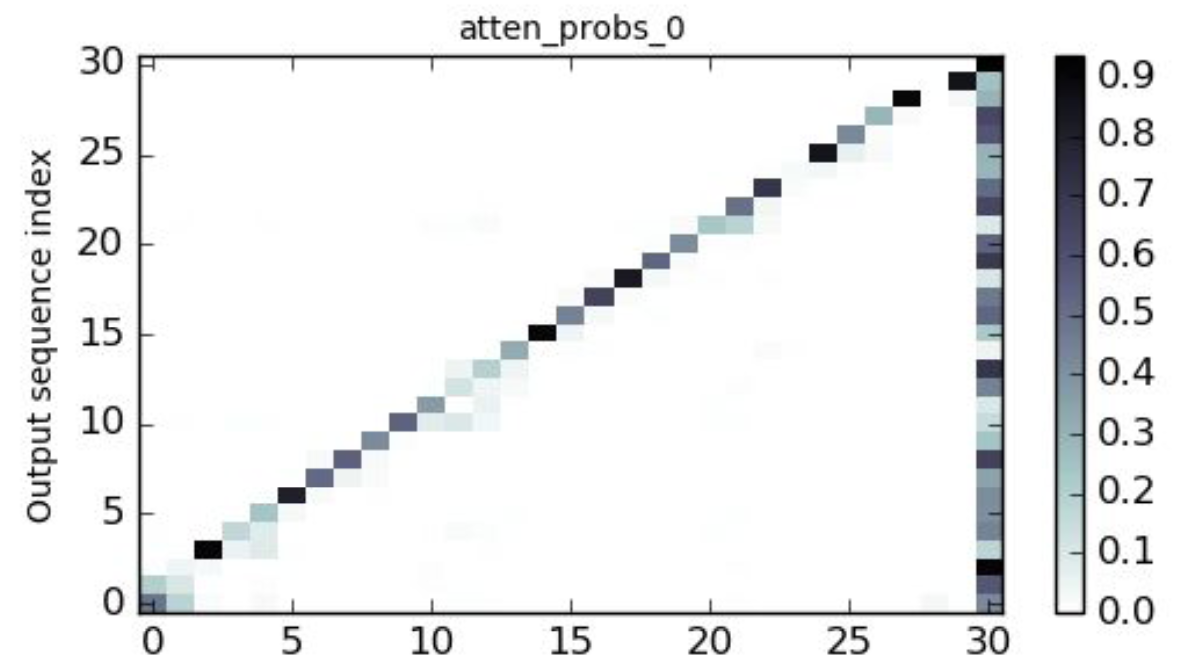}}
\vspace{-2.5mm}
\caption{Example attention weights from the SC model.}
\label{fig:att_weights}
\end{center}
\vspace{-11.0mm}
\end{figure}

First, we calculate performance of the SC method when only considering the top hypothesis output by the recognizer. As shown in row 5 of Table~\ref{table:baseline-text-only}, correcting the top hypothesis gives a 15.8\% relative improvement over the baseline.
Figure~\ref{fig:att_weights} shows example attention weights using in the SC model.  We find that the attention weights are generally monotonic, and where errors occur, the attention weights are aligned to adjacent context, helping the model to choose a more suitable output.  This behavior can be seen between output tokens 10 to 15 in Figure~\ref{fig:att_weights}.

\subsubsection{Generating richer n-best lists}
Table~\ref{table:oracle} compares oracle WERs of the LAS model with and without spelling correction. Two configurations of the SC model are considered: correcting the top hypothesis emitted by the LAS model, leading to final list of 8 candidates, and correcting all 8 entries in the LAS n-best list, leading to an expanded list of 64 candidates.
When correcting only the top LAS hypothesis, the SC model gives only a small improvement in oracle WER.
However, when applying the SC model independently to each entry in the full LAS n-best list, the oracle WER is significantly reduced to almost half.
This demonstrates that the SC model is able to generate a richer and more realistic list of hypotheses, which is more likely to include the correct transcript.
\begin{table}[t]
  \centering
  \begin{tabular}{lcc}
    \toprule
    \textbf{System} & \textbf{Dev-clean} & \textbf{Test-clean} \\
    \midrule
    \text{LAS} & 3.11 & 3.28 \\
    \text{LAS \then SC (1)}   & 3.01 & 3.02 \\
    \text{LAS \then SC (8)}   & \textbf{1.63} & \textbf{1.68} \\
    \bottomrule
  \end{tabular}
  \vspace{-1.0mm}
  \caption{Oracle WER before and after applying the SC model.}
  \label{table:oracle}
\end{table}

This motivates the use of LM rescoring on the expanded n-best list as introduced in Section~\ref{sssec:inc_nbest}. The optimal weights found by tuning on dev set are: $\lambda_{LAS} = 0.7$, $\lambda_{SC} = 1.0$, and $\lambda_{LM} = 0.1$. We use the same parameters on the test set.
As shown in row 6 of Table~\ref{table:baseline-text-only}, this leads to a significant performance improvement compared to using the SC model alone, corresponding to more than a 28\% relative improvement over the baseline LAS model alone.  This promising result shows that the probability scores emitted by each of the three models are complementary to each other.

\subsubsection{Train on more realistic TTS dataset}

Using TTS data to synthesize errors the LAS model makes is not perfect as there is a mismatch between the TTS data and real audio. Thus, when we compare the decoded errors generated from TTS data and real-audio data, as shown in Table \ref{table:cmp tts}, there is still big mismatch between them. Specifically, the table shows the performance of LAS model on LibriSpeech dev set and a TTS dev set generated using the same transcripts. The SC model which is applied on a TTS test set performs better than the real audio test set, even after LM rescoring.

\begin{table}[t]
  \centering
  \begin{tabular}{lcc}
    \toprule
    \textbf{System} & \textbf{Dev-clean} & \textbf{Dev-TTS} \\
    \midrule
    \text{LAS baseline} & 5.80 & 5.26 \\
    \text{LAS \then SC (1)}   & 5.04 & \textbf{3.45} \\
    \text{LAS \then SC (8) \then LM (64)} & 4.20 & \textbf{3.11} \\
    \bottomrule
  \end{tabular}
  \vspace{-1.0mm}
  \caption{WER comparison on a real audio and TTS dev sets.}
  \label{table:cmp tts}
  \vspace{-4.2mm}
\end{table}

To address this audio mismatch issue, we look to add noise to the TTS data to make sound less "clear" and thus have a noisier n-best list. The last two rows in Table \ref{table:baseline-text-only} summarizes the results of SC models trained on errors decoded from MTR TTS data. Performance of the SC model with MTR data improves over clean data. Overall, after applying LM rescoring to the MTR-ed SC model, we achieve  a  29.0\% relative improvement over the LAS baseline.

\subsection{Error analysis}
To understand the errors made by LAS model with LM rescoring before and after spelling  correction, we pulled a few representative examples. Table \ref{table:examples} shows examples of where "LAS + SC + LM rescore" system wins over the "LAS + LM rescore" system. The examples show that SC model does correct many errors of proper nouns and rare words, and also some tense and grammar errors.

\renewcommand\arraystretch{0.85}
\begin{table}[h!]
  \centering
  \begin{tabular}{ll}
    \toprule
    \textbf{LAS + LM rescore} & \textbf{LAS + SC + LM rescore} \\
    \midrule
    \specialcell{ready to hand over to trevellion} & \specialcell{ready to hand over to
\textbf{trevelyan}}\\
    \\
    \specialcell{has countenance the belief the \\hope the wish that the \\epeanites
or at least the \\ nazarines} & \specialcell{has \textbf{countenanced} the belief\\ the hope the wish
that the\\ \textbf{ebionites} or at least the \\ \textbf{nazarenes}}\\
    \\
    \specialcell{a wandering tribe of\\ the blamis
or nubians} & \specialcell{a wandering tribe of\\ the
\textbf{blemmyes} or nubians}\\
    \bottomrule
  \end{tabular}
  \vspace{-1.0mm}
  \caption{LAS + SC + LM rescore Wins. LAS + LM rescore (in bold)}
  \label{table:examples}
  \vspace{-6.0mm}
\end{table}

\section{Conclusions}
In this work we propose a spelling correction model which is trained to explicitly correct the errors made by a speech recognizer. To train the SC model, we generate error hypotheses by decoding the TTS data synthesized from a large text-only corpus. Our results show that the SC model yields clear improvement over the baseline LAS model when directly correcting top LAS hypothesis. By correcting all the entries in the LAS n-best list, the SC model can generate an expanded list which has significantly lower oracle WER. When further rescoring the expanded n-best list with an external LM, it outperforms the simple LM rescoring and direct LAS model training on the TTS data. In order to make further improvement by alleviating the mismatch between TTS data and real audio, we train the SC model on MTR TTS data. Performance of the SC model with MTR data clearly improves over clean data.

\section{Acknowledgements} 
The authors would like to thank Zelin Wu, Anjuli Kannan, Dan Liebling, Rohit Prabhavalkar, Kazuki Irie, Golan Pundak, Melvin Johnson, Mia Chen, Zhouhan Lin, Antonios Anastasopoulos and Uri Alon for helpful discussions.

\bibliographystyle{IEEEbib}

\end{document}